\documentclass[twocolumn,pra,showpacs,superscriptaddress]{revtex4}
\usepackage{graphicx}
\usepackage{amsmath}
\begin{document}

\title{Intrinsic intensity fluctuations in random lasers}
\date{\today}
\author{Karen L. van der Molen}
\email{k.l.vandermolen@utwente.nl} \affiliation{Complex Photonic
Systems, MESA$^+$ Research Institute and Department of Science and
Technology\\ University of Twente, PO Box 217, 7500 AE Enschede,
The Netherlands.}
\author{Allard P. Mosk}
\affiliation{Complex Photonic Systems, MESA$^+$ Research Institute
and Department of Science and Technology\\ University of Twente,
PO Box 217, 7500 AE Enschede, The Netherlands.}
\author{Ad Lagendijk}
\affiliation{Photon Scattering, FOM-institute for Atomic and
Molecular Physics (AMOLF), Kruislaan 407, 1098 SJ Amsterdam, The
Netherlands}

\begin{abstract}
We present a quantitative experimental and theoretical study of
intensity fluctuations in the emitted light of a random laser that
has different realizations of disorder for every pump pulse. A
model that clarifies these intrinsic fluctuations is developed. We
describe the output versus input power graphs of the random laser
with an effective spontaneous emission factor ($\beta$ factor).
\end{abstract}


\maketitle

\section{Introduction}
In 1968 Letokhov wrote his pioneering paper \cite{Letokhov1968} in
which he predicted that light amplification through stimulated
emission is possible in a random medium with gain. A preeminent
experimental demonstration of such a \textit{random laser} was
published by Lawandy and coauthors in 1994.\cite{Lawandy1994}
Typical phenomena of the random laser are: a (low) threshold in
the power conversion, spectral narrowing, and sharp features
("spikes") in the emitted spectrum for both picosecond
\cite{Cao1998, Mujumdar2004,Polson2003,Milner2005} and nanosecond
\cite{Frolov1999} pump pulses. Due to their low threshold and
their ease of production, random lasers are expected to be used in
many applications, such as coding of clothing
\cite{Ramachandran2002} and detection of dangerous materials
\cite{Rose2005}. The theoretical search for the underlying
principles of random laser intensifies, focussing on their
statistical properties and
fluctuations.\cite{Deych2005,Angelani2006,Yamilov2005,Sharma2006}

Recently, a new random-laser phenomenon was described by Anglos
and coauthors \cite{Anglos2004}. They observed shot-to-shot
intensity fluctuations in the emitted light, which were not caused
by fluctuations in the pump source. In their experiments, the
intensity fluctuations occur for nanosecond pump pulses, but not
for picosecond pump pulses. The physical understanding of the
underlying principle of these fluctuations is relevant for
applications and our perception of random lasers.

In this paper we present a quantitative experimental and
theoretical study of the statistics of these shot-to-shot
intensity fluctuations of random laser systems that have different
realizations of disorder for every pump pulse, and develop a model
that clarifies the existence of these fluctuations. Our aim is to
provide physical understanding of the fluctuations. With such a
model, experimental conditions can be tailored to control the
fluctuations.

This paper is organized as follows. In section~\ref{sec,setup} the
experimental setup is described, followed in
section~\ref{sec,measurement} by the experimental observations. A
model based on the number of laser modes in the random laser is
presented in section~\ref{sec,model}. We compare our model with
experimental observations in section~\ref{sec,comparison}.

\section{Apparatus and samples \label{sec,setup}}
We start by describing the experimental details of our samples and
the setup. The random laser consists of a suspension of TiO$_2$
particles (mean diameter of 180 nm) in a solution of methanol and
Sulforhodamine B (1 mmol/liter). The suspension is contained in a
fused silica capillary tube, with internal dimensions 100 $\times$
2 $\times$ 0.2 mm$^3$. To characterize the mean free path of this
sample, we performed an enhanced-backscatter cone experiment
\cite{Albada1985} and an escape function experiment
\cite{Gomez2003}. We found a transport mean free path of 0.46
$\pm$ 0.1 $\mu$m at $\lambda$ = 633 nm.

The samples are excited by a pump pulse at 532 nm, provided by an
optical parametric oscillator (OPO) pumped by a Q-switched Nd:YAG
laser (Coherent Infinity 40-100/XPO). The pump
pulse has a duration of 3 ns and a repetition rate of 50 Hz. 
The pump light is focused with a microscope objective
(water-immersed, NA = 1.2) onto the sample (focus area = 12 $\pm$
6 $\mu$m$^2$), reaching an intensity in the order of 1 mJ/mm$^2$.
The light emitted by the random laser is collected by the same
objective. A small part of the pump light is split off from the
input beam before the objective. This light is used to excite a
dye solution (Cresyl Violet) which we use as a marker for the pump
fluence. This dye solution works as a wavelength converter for the
pump light. The light emitted by Cresyl Violet and by the random
laser are recorded at the same time with a spectrometer and an
intensified charged coupled device (resolution $\sim$ 0.3 nm
spectral width): in one frame both the emission of the random
laser and the pump marker is recorded. The pump light is filtered
out of the detection path by use of a colored glass filter with a
transmission of less than 1\% at the wavelength of the pump laser.

As a picosecond source we use a stretched pulse at 532 nm of a
femtosecond optical parametric amplifier (OPA), pumped by a
femtosecond Ti:Sapphire laser. We could not measure the pump pulse
duration directly, but calculated a duration of 15 $\pm$ 0.5 ps
from the measured bandwidth of the OPA and the configuration of
the pulse stretcher. Measurements of the random laser and Cresyl
Violet simultaneously were not performed with the picosecond pump
pulse.

\section{Measured fluctuations \label{sec,measurement}}
We are interested in the intrinsic fluctuations, i.e. the
fluctuations of the random laser itself which are not the result
of fluctuations of the pump laser. A typical single shot emission
spectrum shows two peaks, see the inset of
Fig.~\ref{fig,randomlaserfluctuations}: the peak of the random
laser output is around 594 nm, and the peak of the marker Cresyl
Violet is around 625 nm. The intrinsic fluctuations of the random
laser are investigated by comparing the peak height of the light
emitted by the random laser with that of the marker. We have taken
many single shot measurements ($>$ 400), and from each of these
spectra we determine the two peak heights. The data is shown in
Fig.~\ref{fig,randomlaserfluctuations}. The corresponding
correlation coefficient is 0.4, indicating that a large part of
the fluctuations is intrinsic. In contrast, we expect stronger
correlation between the spontaneous emission regime of the random
laser and the pump fluence. To check this expectation we compute
the correlation coefficient for the spectral radiance at 573 nm,
relatively far away from the random laser emission peak, and the
pump marker. For this situation we find a stronger correlation
coefficient of 0.62, indicating that these fluctuations are mainly
due to the pump fluctuations. Although the spectral radiance at
573 nm is still partially influenced by stimulated emission, the
effect of the spontaneous emission can already be seen from the
increase of the correlation coefficient.

\begin{figure}
 \includegraphics[width=3.2in]{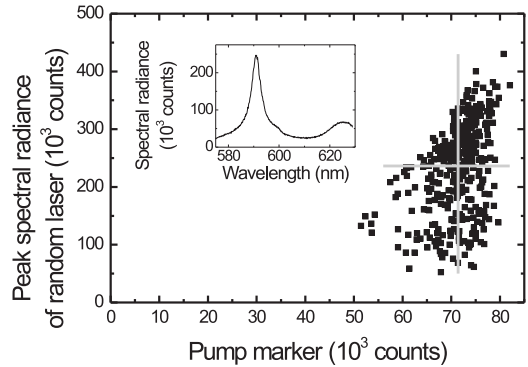}
\caption{\label{fig,randomlaserfluctuations}The peak of the
spectral radiance of the random laser is plotted versus the pump
marker. The two gray lines indicate the mean values of the two
peak heights and are used to visualize the correlation. The inset
shows a typical single shot spectrum, from which the counts of the
peak heights are determined.}
\end{figure}

The intrinsic fluctuation coefficient $f$ is defined as the ratio
between the standard deviation of the shot-to-shot intensity of
the light emitted by the random laser, $\Delta I$, and the mean
value of the intensity of the emitted light $\overline{I}$.
\begin{eqnarray}
f \equiv \frac{\Delta I}{\overline{I}}
\end{eqnarray}
To determine the intrinsic fluctuations from the data points in
Fig.~\ref{fig,randomlaserfluctuations} we take a small band of
pump fluences so that the fluctuations of the pump laser do not
influence the outcome. We make a histogram of the accompanying
peak heights of the spectral radiance of the random laser and fit
this histogram with a gaussian. From this fit we obtain the
standard deviation and the mean value of the peak heights, leading
us to the experimental fluctuations
\begin{eqnarray}
f_{\rm{ns,exp}} = (18 \pm 3)\%,
\end{eqnarray}
where the error margin corresponds to one standard deviation.

Due to large technical fluctuations of the pump laser, no reliable
estimate of the intrinsic fluctuations follows from the picosecond
pumped random laser experiments.

\section{Model \label{sec,model}}
In this section we present a model for the origin of the intrinsic
fluctuations of random lasers. The approach we take here is based
on the concept of pseudo modes~\cite{Ling2001}. Pseudo modes are
single frequency eigen modes, solutions of Maxwell equations.
These modes have an eigenfrequency and decay by leaking to the
outside world. This leakage is characterized by a decay time
$\tau_c$. Each mode can be a laser mode, depending on the decay
time (also referred to as dwell time) of that mode and the gain
time of the sample. The decay time is the time that light is
inside the sample due to diffusion, and the amplitude gain time is
defined as the time after which the amplitude is increased with a
factor $e$. If the decay time of a certain pseudo mode is longer
than the gain time of the system, that pseudo mode is a laser
mode. We assume the mode volume is equal for each (random) mode,
and that the gain in the sample is homogeneous. The number of
lasing modes, $N_l$, is given by
\begin{eqnarray}
N_l = \pi_l N, \label{eqn,modelasing}
\end{eqnarray}
with $N$ the number of pseudo modes and $\pi_l$ is a random
variable ranging between 0 and 1. We define
\begin{eqnarray}
p_l \equiv \overline{\pi_l} , \label{eqn,defpl}
\end{eqnarray}
with $p_l$ the probability for lasing in a pseudo mode, and
$\overline{\pi_l}$ is $\pi_l$ averaged over realizations of the
disorder.The emission power of the different lasing modes is
equal. The detected emission intensity of the random laser is
almost entirely due to the lasing modes in the system and is
assumed to be proportional to the number of lasing modes. The
intrinsic fluctuations $f$ can be determined
\begin{eqnarray}
f = \frac{\Delta I}{\overline{I}} =
\frac{\sqrt{\overline{N_l}}}{\overline{N_l}}. \label{eqn,modefluc}
\end{eqnarray}
In the last step of Eq.~(\ref{eqn,modefluc}), we have assumed a
binomial distribution of $N_l$, which results for the limit of $N$
to infinity to a gaussian or normal distribution. The standard
deviation, $\sigma$, is given by $\sqrt{\overline{N_l}}$. We
combine Eqs.~(\ref{eqn,modelasing}) and~(\ref{eqn,defpl}), and
insert the result in Eq.~(\ref{eqn,modefluc}) to obtain
\begin{eqnarray}
f = \frac{1}{\sqrt{p_l N}}. \label{eqn,modeflucplN}
\end{eqnarray}

In section~\ref{sec,probabilityoflasingtheory} we will show how to
calculate $p_l$ from a fit to experimental data. An elaboration on
the calculation of $N$ is presented in
section~\ref{sec,numberofmodes}.

\subsection{Determination of the probability of lasing
\label{sec,probabilityoflasingtheory}} When we combine
Eqs.~(\ref{eqn,modelasing}) and~(\ref{eqn,defpl}) the probability
of lasing is given by
\begin{eqnarray}
p_l = \frac{\overline{N_l}}{N}. \label{eqn,defplNlN}
\end{eqnarray}
The probability of lasing can be calculated via the distribution
of the decay times. The integral of the distribution of the decay
times $P$ from the gain time to infinity will give the probability
$p_l$ for a certain mode to be a laser mode:
\begin{eqnarray}
p_l = \int_{\tau_g}^{\infty} P(\tau_c) d\tau_c. \label{eqn,intpl}
\end{eqnarray}
The distribution of the decay times for a 3D diffusive medium is
not known. We therefore use the distribution of the phase delay
times \cite{Tiggelen1999}, which is expected to be close to the
distribution of the decay times.

The gain time is given by
\begin{eqnarray}
\tau_g \equiv \frac{\ell_g n'}{c_0},
\end{eqnarray}
where $n'$ is the real part of the refractive index. The amplitude
gain length $\ell_g$ is given by
\begin{eqnarray}
\ell_g = \frac{2}{\sigma_e \rho_{\rm{exc}}},
\label{eqn,gainlength}
\end{eqnarray}
where $\sigma_e$ is the stimulated emission cross section of a
molecule, and $\rho_{\rm{exc}}$ is the density of molecules in the
excited state in the sample.

If the pump power is large enough, the gain in the system can be
saturated. In the case of saturation the gain length will not be
decreased any more when the pump power is increased. From
Eq.~(\ref{eqn,gainlength}) we can find a lower bound for the gain
length $\ell_{g,b}$, and thus an indication of gain saturation,
when one assumes that all the dye molecules in the medium are in
the excited state
\begin{eqnarray}
\ell_{\rm{g,b}} \geq \frac{2}{\sigma_e \rho},
\label{eqn,gainlengthlowerbound}
\end{eqnarray}
where $\rho$ is the density of dye molecules in the sample.

We want to determine the probability of lasing directly from our
experiments. When we examine Eq.~(\ref{eqn,defplNlN}), we see a
similarity between the definition of $p_l$ and the spontaneous
emission factor of a laser, the $\beta$ factor
\cite{Woerdman2001a}. The singlemode $\beta$ factor, defined as
the fraction of spontaneous emission that contributes to lasing is
given by \cite{Siegman1986}
\begin{eqnarray}
\beta_{\rm{sm}} = \frac{1}{N}.
\end{eqnarray}
This $\beta_{\rm{sm}}$ factor appears in the four-level rate
equations for a singlemode laser \cite{Siegman1986}
\begin{subequations}\label{eqn,rate}
\begin{eqnarray}
\frac{dN_1(t)}{dt} & = & P_L(t) - \frac{\beta_{\rm{sm}}
q(t)N_1(t)}{\tau}-\frac{N_1(t)}{\tau}, \label{eqn,rateN1}
\\
\frac{dq(t)}{dt} & = & - \frac{q(t)}{\tau_c}
+\frac{\beta_{\rm{sm}} N_1(t)}{\tau}\left( q(t) + 1 \right),
\label{eqn,rateq}
\end{eqnarray}
\end{subequations}
with $N_1$ the number of excited molecules in the medium, $q$ the
number of photons in the lasing mode, $P_L$ the pump rate (in
photons per second), $\tau$ the spontaneous emission life time of
the dye and $\tau_c$ the cavity decay time.

In a random laser many random modes contribute to the laser
oscillation. However, for our consideration only the average
behavior is relevant. In general, to describe a multimode laser
one has to write an equation for every mode and couple the
different mode equations to the equation for the population. Only
two small changes to the singlemode rate equations are necessary,
if the behavior of the multimode laser can be simulated by the
behavior of a singlemode laser. We simply replace
$\beta_{\rm{sm}}$ and $\tau_c$ in the rate
equations~(\ref{eqn,rate}) by the effective parameters
$\beta_{\rm{mm}}$ and $\tau_{\rm{c,mm}}$. Since we are interested
in the average behavior, we will use this simplified approach. We
will prove that this approach is valid in many situations.

For the mean value of the cavity decay time, given by
$\tau_{\rm{c,mm}}$, we use the mean value of the distribution of
the decay times. To determine this distribution we calculate the
solution of the diffusion equation for a slab with thickness $L$,
with a source positioned in the middle of the slab. From this
solution of the diffusion equation the electric field correlation
is derived and the mean value of the phase delay times follows
from the Taylor expansion of this correlation.\cite{Genack1999}
The mean value of the phase delay time, and thus the mean value of
the cavity decay time, is given by
\begin{eqnarray}
\tau_{\rm{c,mm}} = \frac{1}{8} \frac{L^2}{D}. \label{eqn,tauc}
\end{eqnarray}
The diffusion constant $D$ is given by $c_0 \ell / (3n')$, with
$\ell$ the transport mean free path.\cite{Sheng1995} For the
effective parameter $\beta_{\rm{mm}}$ we take
\begin{eqnarray}
\beta_{\rm{mm}} = \frac{\overline{N_l}}{N}, \label{eqn,defbmm}
\end{eqnarray}
where we assumed that all modes contribute equally. Comparing
Eqs.~(\ref{eqn,defplNlN}) and~(\ref{eqn,defbmm}) we see that
$\beta_{\rm{mm}}$ is equal to $p_l$.

In the continuous wave limit for a singlemode laser in steady
state a formula can be analytically derived from the rate
equations~(\ref{eqn,rate}) that describes the relation between the
output and the input power of a laser.
\begin{eqnarray}
q = \frac{1}{2} \sqrt{ \left( \frac{1 - P_L \tau_{c}
\beta_{\rm{sm}} }{\beta_{\rm{sm}}} \right )
^2+4P_L\tau_{c}}-\frac{1 - P_L \beta_{\rm{sm}} \tau_{c}}{2
\beta_{\rm{sm}} }. \label{eqn,fitbeta}
\end{eqnarray}
If one uses a pulsed pump this should be generalized to include
time dependence, and an analytic solution is no longer available.
However, as we will show, for a wide range of parameters
Eq.~(\ref{eqn,fitbeta}) still describes the threshold behavior
very well, even for a detector that integrates the output power.
We will use Eq.~(\ref{eqn,fitbeta}) for the integrated power of a
multimode laser, with the replacements of the parameters as
described above. For the use of Eq.~(\ref{eqn,fitbeta}) in pulsed
experiments we replace the parameter $\beta_{\rm{mm}}$ by an
effective parameter $\beta_{\rm{eff}}$.

To examine the applicability of Eq.~(\ref{eqn,fitbeta}) to
experiments with a pulsed pump, we calculate with the rate
equations~(\ref{eqn,rate}) several output versus input power
graphs. We use input parameters relevant to our experiment and
vary the pump pulse duration and $\beta_{\rm{sm}}$. The pump pulse
is modelled by a gaussian. To the output versus input power graphs
we fit Eq.~(\ref{eqn,fitbeta}) and use as fit parameter
$\beta_{\rm{eff}}$. In Fig.~\ref{fig,betafactor} we present the
calculated $\beta_{\rm{eff}}$ as function of $\beta_{\rm{mm}}$.
For a pulse duration $t_p$ of 3000 ps, the calculated values of
$\beta_{\rm{eff}}$ are identical to the input value of
$\beta_{\rm{mm}}$. This correspondence implies that for our system
nanosecond pump pulses can be treated as a continuous wave pump.
When the pulse duration is shorter, while keeping the other
parameters constant, we notice a deviation from this straight line
to lower values of $\beta_{\rm{eff}}$ for the same values of
$\beta_{\rm{mm}}$. This difference between $\beta_{\rm{eff}}$ and
$\beta_{\rm{mm}}$ increases for shorter pulses. The origin of the
dissimilarity between $\beta_{\rm{mm}}$ and $\beta_{\rm{eff}}$ is
due to the fact that the pump is not a continuous wave, but a
pulse with a finite duration.

\begin{figure}
\begin{center}
\includegraphics[width=3.2in]{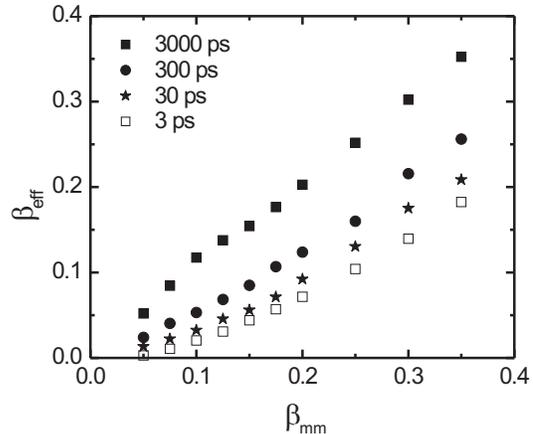}
\end{center}
\caption{\label{fig,betafactor} The theoretical values of the
effective $\beta$ factor, $\beta_{\rm{eff}}$, numerically
calculated from the rate equations, versus the input parameter
$\beta_{\rm{mm}}$ for different pump pulse durations. We choose
the input parameters relevant for our experiments: $\tau$ = 3200
ps and $\tau_c$ = 0.1 ps. In the legend the pump pulse durations
are listed.}
\end{figure}

Besides the pump pulse duration and the spontaneous life time,
there is a third time scale in the rate equations: the mean cavity
decay time. If the pump pulse duration approaches the value of the
cavity decay time $\tau_{\rm{c,mm}}$ Eq.~(\ref{eqn,fitbeta}) is no
longer a good fit to the output versus input power graph. This
failure of the fit means that $\beta_{\rm{eff}}$ is no longer a
parameter that can be used to describe the experiment and thus
that for that experiment there is no $\beta$ factor. In short:
\begin{eqnarray}
t_p & \geq \tau \gg \tau_c & \quad \quad \textrm{CW limit: }\beta_{\rm{eff}} = \beta_{\rm{mm}}, \nonumber \\
\tau & > t_p > \tau_c & \quad \quad \textrm{conversion needed: }\beta_{\rm{eff}} < \beta_{\rm{mm}}, \nonumber \\
\tau & > \tau_c > t_p & \quad \quad \textrm{simplified model
fails}. \nonumber
\end{eqnarray}

We conclude that the rate equations and the threshold
curve~(\ref{eqn,fitbeta}) can be used for a multimode laser, and
that these formulas can be used under certain conditions for a
pumped laser. 
This allows us to extract $p_l$ from the threshold curve of a
random laser.

\subsection{Calculation of the number of modes \label{sec,numberofmodes}}
The second important parameter to calculate the intrinsic
fluctuations is $N$. The total number of modes in the system
within the relevant frequency bandwidth $\Delta \omega$, can be
calculated using the formula~\cite{Siegman1986}
\begin{eqnarray}
N & = & \rho(\omega ,V) \frac{ \Delta \omega}{\omega},
\label{eqn,modenumbera} \\
 & = & \frac{8 \pi n^3 V}{\lambda_c^3} \frac{\Delta
\lambda}{\lambda_c}. \label{eqn,modeamount}
\end{eqnarray}
with  $\rho(\omega ,V)$ the density of field modes in the cavity
volume $V$, $\Delta \lambda$ the full width at half maximum of the
emission spectrum, and $\lambda_c$ the central wavelength of the
emission spectrum. All parameters can be deduced from experiments,
except the volume of the cavity $V$. In our case, the absorption
length is much larger than the transport mean free path, and we
can assume a gain volume in the form of hemisphere
\begin{eqnarray}
V = \frac{2}{3}\pi r^3, \label{eqn,modevolume}
\end{eqnarray}
with $r$ the radius of the gain volume.

The intrinsic fluctuations depend on the number of modes, and the
probability of lasing. By experimentally varying the quantities of
the parameters, the intrinsic fluctuations of the random laser can
be controlled.

\section{Results \label{sec,comparison}}
We will now calculate with our model the intrinsic intensity
fluctuations of a random laser system pumped with nanosecond or
picosecond pulses, and compare these fluctuations with
experimental observations.

\subsection{Probability of lasing \label{sec,probabilityoflasing}}
We have measured the peak of the spectral radiance of the random
laser as a function of the pump fluence, as shown in
Fig.~\ref{fig,outputinputpowerns}. In case of nanosecond pump
pulses we have seen that $p_l$ = $\beta_{\rm{eff}}$. From the fit
of Eq.~(\ref{eqn,fitbeta}) to the experimental data we find
directly the probability of lasing
\begin{eqnarray}
p_{\rm{l,ns}} = 0.07 \pm 0.03,
\end{eqnarray}
where the error margin corresponds to one standard deviation.

\begin{figure}
\begin{center}
\includegraphics[width=3.2in]{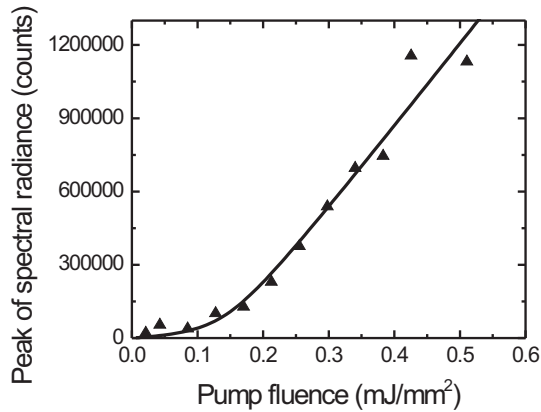}
\end{center}
\caption{\label{fig,outputinputpowerns} Peak of the measured
spectral radiance of the random laser versus the pump fluence for
a random laser pumped with nanosecond pulses. A normal threshold
behavior is observed. The solid line is a fit to our data, with
$\tau_{\rm{c,mm}} = 0.1$ ps, and the spontaneous emission life
time is 3200 ps. From the fit we obtain $\beta_{\rm{eff}}$ = 0.07,
which in our model equals the probability of lasing $p_l$.
[Eq.~(\ref{eqn,fitbeta})]}
\end{figure}

For the picosecond pumped random laser we have also measured the
peak of the spectral radiance of the random laser as a function of
the pump fluence, see Fig.~\ref{fig,outputinputpowerps}. The
parameter $\beta_{\rm{eff}}$ has to be converted to
$\beta_{\rm{mm}}$ ($p_l$). In our picosecond experiment the pulse
duration is $15 \pm 0.5$ ps and a relevant conversion graph for
$\beta_{\rm{mm}}$ and $\beta_{\rm{eff}}$ is presented in
Fig.~\ref{fig,betafactorps}. With this graph we convert our values
of $\beta_{\rm{eff}}$ of 0.03 $\pm$ 0.006 to
\begin{eqnarray}
p_{\rm{l,ps}} & = & 0.09 \pm 0.015,
\end{eqnarray}
where the error margin corresponds to one standard deviation.

\begin{figure}
\begin{center}
\includegraphics[width=3.1in]{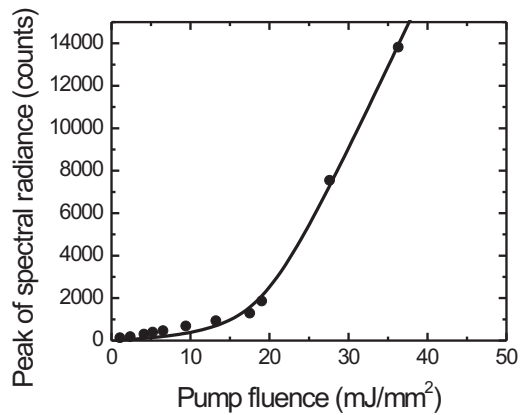}
\end{center}
\caption{\label{fig,outputinputpowerps} Peak of the measured
spectral radiance versus pump fluence for a random laser pumped
with picosecond pulses. The solid line is a fit to our data, with
the parameters $\tau$ = 3200 ps and the mean cavity decay time 0.1
ps. From the fit we obtain $\beta_{\rm{eff}}$ = 0.03.
[Eq.~(\ref{eqn,fitbeta})]}
\end{figure}

\begin{figure}
\begin{center}
\includegraphics[width=3.55in]{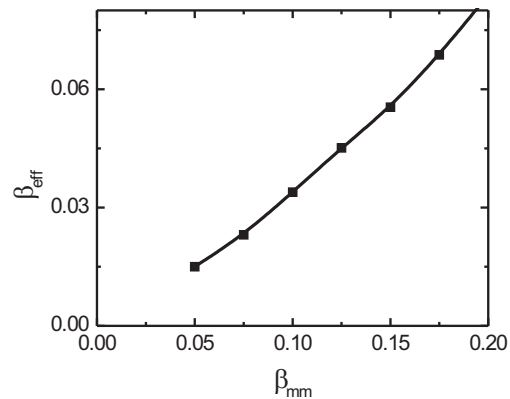}
\end{center}
\caption{\label{fig,betafactorps} The calculated
$\beta_{\rm{eff}}$ versus the input parameter of the rate
equations $\beta_{\rm{mm}}$ is shown. The graph is produced for a
pump pulse duration of 15 ps, $\tau_{\rm{c,mm}}$ = 0.1 ps and
$\tau$ = 3200 ps. The solid lines connect the data points. We
convert our $\beta_{\rm{eff}}$ of 0.03 to a $p_l$ of 0.09.}
\end{figure}

\subsection{Number of modes}
The calculation of the number of modes for our random laser regime
is given by Eq.~(\ref{eqn,modeamount}). In our case we have
$n=1.4837$, $\lambda_c = 595$ nm. The width of the emission
spectrum above threshold is 4.3 nm. The volume of the gain medium
is given by the volume of a hemisphere, see
Eq.~(\ref{eqn,modevolume}). The spatial form of the luminescence
coming from the surface of the random laser sample pumped with
nanosecond pump pulses was recorded with a charge coupled device,
while we filter the pump light. We measure a circular spot with a
mean radius of 5 $\pm$ 0.5 $\mu$m. The number of modes in the
nanosecond pumped situation is
\begin{eqnarray}
N_{ns} = 746 \pm 256.
\end{eqnarray}

We did not record the spatial form of the luminescence in case of
picosecond pump pulses. However we can speculate about the gain
volume in the picosecond pumped system if we have gain saturation.
For our random laser, the probability of lasing for the picosecond
and for the nanosecond pumped case are within each other error
margins, an indication of gain saturation which we shall prove
now.

We calculate the lower bound of the gain length with
Eq.~(\ref{eqn,gainlengthlowerbound}). The density of molecules in
the random laser is 5.4 $\times$ 10$^{23}$ molecules m$^{-3}$, and
the stimulated emission cross section is 4 $\times$ 10$^{-20}$
m$^{2}$, both with a 10\% error, leading to a lower bound limit of
\begin{eqnarray}
\ell_{\rm{g,b}} \geq 75 \textrm{ $\mu$m}.
\end{eqnarray}
For the nanosecond pumped random laser we found a $p_l$ of 0.07
$\pm$ 0.03, see Fig.~\ref{fig,outputinputpowerns}. The gain length
at which $p_l$ equals 0.04 is
\begin{eqnarray}
\ell_{g,b,c} = 60 \textrm{ $\mu$m}.
\end{eqnarray}
The value of $\ell_{g,b,c}$ is lower than the bound limit we
calculated. Although this does not indicate a serious discrepancy
as the error margins on $p_l$ are only one standard deviation, the
assumption that the phase delay time distribution equals the
cavity decay time distribution should be investigated. Both gain
lengths above are almost equal, proving that we are near the
saturation regime. The threshold pump fluence is a factor 100
higher in the case of the picosecond pump pulse, and the pulse
duration of the picosecond pump pulse is a factor 100 shorter.
Since both pump lasers pump the sample to gain saturation, the
gain volume of the picosecond pump pulse is larger than the gain
volume of the nanosecond pump pulse. The advantage of using two
different pump lasers on one sample, is the change in the number
of laser modes.

The gain volume for the picosecond case in saturation is at
minimum a cylinder with a radius equal to the luminescence spot of
the nanosecond pumped random laser and a length equal to $L_d$.
The length $L_d$ is the length that light travels from a point
source inside a diffusive medium
\begin{eqnarray}
L_d = \sqrt{D t}, \label{eqn,lengthLd}
\end{eqnarray}
with $t$ the pulse duration of the point source. In our system
$L_d$ is 22 $\pm$ 3 $\mu$m. This leads to a total number of modes
in the picosecond pumped random laser of
\begin{eqnarray}
N_{ps} \geq 2407 \pm 885.
\end{eqnarray}

\subsection{Intrinsic intensity fluctuations of a random laser}
The intensity fluctuations derived from our model are given by
Eq.~(\ref{eqn,modefluc})
\begin{eqnarray}
f_{ns} = (14 \pm 5) \%,
\end{eqnarray}
where the error margin corresponds to one standard deviation. This
number for the fluctuations is in good agreement with our
experimental observations of (18 $\pm$ 3)\%.

The intrinsic fluctuations calculated for a picosecond pump pulse
for our own system is
\begin{eqnarray}
f_{ps} \leq (6.8 \pm 1.2) \%,
\end{eqnarray}
where the error margin corresponds to one standard deviation. We
could not verify this result experimentally. Anglos and coauthors
have performed measurements on a random laser pumped with
picosecond pulses.\cite{Anglos2004} From their paper we can
calculate the fluctuations of their system with our model. From
the fit to the published output versus input power graph, and with
a conversion from $\beta_{\rm{eff}}$ to $p_l$,  we find for their
system $p_l$ = 0.05. For the gain volume we assumed a cylindrical
form, as their excitation spot is much larger than their mean free
path. We obtain $f_{ps}$ = 0.01\%. From their published measured
spectra we can find an
estimation for the measured fluctuations. 
The measured fluctuations are (3.7 $\pm$ 1.7)\%, where the error
margin corresponds to one standard deviation. These fluctuations
include the pump fluctuations. These pump fluctuations can be in
the order of 3\% for a typical picosecond laser source. Since the
expected intrinsic fluctuations are much smaller than the pump
fluctuations, the intrinsic fluctuations cannot be measured and
our model is not inconsistent with their data.

\section{Conclusions}
We have developed a model based on quasi modes that predicts the
fluctuations of the output power of a random laser pumped with
either nanosecond or picosecond pulses. For the system pumped with
nanosecond pulses we computed fluctuations of $(14 \pm 5)$\%. This
is in good agreement with our experimental fluctuations of $(18
\pm 3)$\%. For a system pumped with picosecond pulses we
calculated $f_{ps}$ = 0.01\%, too small to be observed, for the
system measured by Anglos and coauthors \cite{Anglos2004}.

The difference in intrinsic fluctuations between picosecond and
nanosecond pumped random lasers is well described by our model and
the predictions are identical to our observations for a nanosecond
pumped random laser and published observations for a picosecond
pumped random laser. Our model can be used to tailor experimental
conditions in such a way as to control the intrinsic fluctuations
of a random laser system.

\section*{Acknowledgements}
We thank Boris Bret for the discussion that lead to the estimation
of the transport mean free path, Tom Savels for fruitful
discussions considering the model, and Tijmen Euser and Willem Vos
for use of the femtosecond OPA system. This work is part of the
research program of the 'Stichting voor Fundamenteel Onderzoek der
Materie' (FOM), which is financially supported by the 'Nederlandse
Organisatie voor Wetenschappelijk Onderzoek' (NWO).


\end{document}